# Self-Resonant µ-Lasers of Colloidal Quantum Wells Constructed by Direct Deep-Patterning

*Negar Gheshlaghi, Sina Foroutan-Barenji, Onur Erdem, Yemliha Altintas, Farzan Shabani, Muhammad Hamza Humayun, and Hilmi Volkan Demir\**

Dr. N. G., S. F., Dr. O. E., Dr. Y. A., F. S., M. H. H., Prof. H. V. D.
Department of Electrical and Electronics Engineering
Department of Physics
UNAM – Institute of Materials Science and Nanotechnology
Bilkent University
Ankara 06800, Turkey

Dr. Y. A.
Department of Materials Science and Nanotechnology
Abdullah Gul University
Kayseri 38080, Turkey

Prof. H. V. D.
LUMINOUS! Centre of Excellence for Semiconductor Lighting and Displays
Centre of Optical Fiber Technology
The Photonics Institute
School of Electrical and Electronic Engineering
School of Physical and Mathematical Sciences
Nanyang Technological University
50 Nanyang Avenue, Singapore 639798, Singapore
\*E-mail: hvdemir@ntu.edu.sg, volkan@bilkent.edu.tr

Keywords: semiconductor nanocrystals, direct nanopatterning, UV-induced ligand exchange, electron beam lithography, microlaser, optical nanocircuit, colloidal quantum wells

Here, the first account of self-resonant fully-colloidal µ-lasers made from colloidal quantum well (CQW) solution is reported. A deep patterning technique is developed to fabricate well-defined high aspect-ratio on-chip CQW resonators made of grating waveguides and in-plane reflectors. CQWs of the patterned layers are closed-packed with sharp edges and residual-free lifted-off surfaces. Additionally, the method is successfully applied to various nanoparticles including colloidal quantum dots and metal nanoparticles. It is observed that the patterning process does not affect the nanocrystals (NCs) immobilized in the attained patterns and different physical and chemical properties of the NCs remain pristine. Thanks to capability of the proposed patterning method, patterns of NCs with sub-wavelength lateral feature size and micron-scale height are fabricated in the aspect ratios of 1:15 (<100 nm lateral patterned features to >1.5 µm film thickness). The fabricated waveguide-coupled laser, enabling tight optical confinement, assures in-plane lasing. The spectral characteristics of the designed CQW resonator structure are well supported with a numerical model of full electromagnetic solutions. Such directly deep-patterned self-resonant µ-lasers of CQWs hold great promise for on-chip integration to photonic circuits.

## 1. Introduction

Quantum photonics holds great promise for future technologies such as secure communication, quantum computation, quantum simulation, and quantum metrology. The technologies of semiconductor integrated circuits and electronic devices are rapidly approaching their fundamental limits in terms of both processing speed and data transmission rate. One way to overcome this limitation is to employ photons rather than electrons in the functional processing components.[1,2] Although tremendous progress has been made towards the development of scalable miniature circuits, integration of a versatile optical source that couples light into the waveguides as one of the main elements of a photonic circuit is highly desirable because it would avoid the limitations of otherwise delivering light from a bulky external source to the nanometer scale. Micro-lasers (µ-lasers) may provide a viable option as they perform single-mode photon emission and allow for the generation of coherent radiation within an extremely small footprint. However, despite many demonstrations of this class of lasers based on III–V materials and silicon, the required high growth temperatures and large lattice mismatch between laser material and substrate in monolithic integration as well as the difficulties in precise surface flatness required for bonding fabricated lasers on a substrate, and the realization of low cryogenic operation temperature of these lasers present a formidable array of challenges to current fabrication technologies.[3]

Colloidal semiconductor nanocrystals (NCs) are emerging as new optoelectronic materials that have recently been commanding considerable attention in photonics thanks to their size-tunable bandgaps, strong light absorption, narrow spectral emission, chemical stability, and easy processibility.[4-6] Compared to the epitaxial growth of III–V quantum wells, solution-processing of NCs offers low-cost manufacturability in optoelectronics. A variety of types of colloidal quantum dots (QDs) and their two-dimensional counterparts, colloidal quantum wells (CQWs), have been widely explored as optical gain medium and a number of successful optical feedback configurations including Fabry−Pérot cavity,[7,8] whispering gallery mode (WGM),[9,10] distributed feedback (DFB)[11,12] lasers have been developed recently. However, these reports mostly focused on the performance of individual devices, and despite the confirmation of NCs as optical-gain media in these structures, a practical and fundamental way to develop a fully integrated on-chip waveguide-coupled µ-laser sources has yet to be achieved. A challenge in the NC solution-based device is to create high-density close-packed well-defined colloidal nanopatterns with controllable film thickness. Different methods of NC film patterning have been considered and a number of successful NC patterning techniques including UV/photo-curing patterning,[13-16] micro-contact printing[17,18] and patterning through electrospinning[19] have been studied. Still, in most of these methods, colloidally synthesized NCs capped with long-chain organic ligands are patterned directly or blended with organic polymers. Unfortunately, these organic ligands and additives adversely affect device performance. These act as insulating barriers that impede charge or heat transport between NCs and decrease film density, because of which they do not form close-packed films, leading to limited pattern thickness and resolution. These issues become more critical when NCs are used as gain material. The gain value is dependent on the packing density of the NCs inside the surrounding matrix. Low packing density reduces the gain coefficient as well as refractive index of these layers and then their optical mode confinement.[20]

Here, we demonstrate the first account of on-chip self-resonant all-NC µ-laser that incorporates patterned NCs as a gain medium between ultra-compact patterned NCs as mirrors.

The proposed structure has the potential to be coupled to a waveguide to be used in chip-scale photonic devices. Unlike exploitation of simple fabrication methods[21,22] such as "coffee-stain" cavities[23-25], which have resulted in limited control over the cavity geometry, our proposed fabrication approach enables tight confinement of light, significantly enhances the directionality of the output beam and assures in-plane laser emission propagation. Here, the devised fully-colloidal laser architecture relies on self-resonating distributed Bragg reflector (DBR) grating waveguide as the gain section integrated in plane with a pair of reflectors, one of which is a highly reflective (100%) DBR grating while the other is defined as a moderately reflective (90%) DBR outcoupler. All these parts are directly patterned using the same NCs, a specific type of CQWs synthesized in core/hot-injection shell (HIS) both as gain and dielectric media. Indeed, the fabrication of such high-quality resonators requires close-packed, high-contrast and contamination-free subwavelength patterning technique.

In this study we developed a novel deep-patterning method to process NC films in close packing. This technique combines electron beam lithography with UV-induced ligand exchange and enables fabrication of the nano-patterned NC structures. The original long bulky insulating ligands are replaced with shorter compact ones to obtain close-packed patterned films, while intrinsic physical and chemical properties of the NCs are not altered. Using the proposed direct deep-patterning technique, high aspect-ratio NC patterns are fabricated. Lateral feature sizes below 100 nm resolution, limited by the available electron beam lithography (EBL) apparatus are achieved, while feature thicknesses up to 1.5 µm were reproducibly obtained, yielding an aspect ratio of 1:15. These achievable dimensions conveniently allow for construction of fully-colloidal self-resonant µ-laser with grating waveguides as gain medium as well as a pair of in-plane mirrors made of DBRs.

From the fabricated µ-laser, the lasing threshold as low as 21 µJ/cm$^2$ was achieved under femtosecond optical pumping. The lasing action was characterized via measuring the photoluminescence (PL) intensity versus pump power. The achieved lasing threshold is comparable to those obtained using external DFB and WGM schemes with NCs.[11,26-30] In the case of using such an on-chip self-resonant µ-laser cavity, the high refractive index contrast between the patterned CQWs and the air increases the optical mode confinement in the gain medium, which relatively makes it easier to lase. Low-threshold lasing is indicative of efficient in-plane feedback and low-loss waveguiding, therefore a further confirmation of the high structural fidelity of our deep-patterned films.

## 2. Results and Discussion

The fabrication of self-resonator µ-laser started with the patterning of CdSe/Cd$_{0.25}$Zn$_{0.75}$S core/hot-injection shell (HIS) CQWs, which is synthesized using modifications according to the synthesis recipe reported in our group's recent studies [31,32] (see Methods) and followed by conventional lift-off process. To assemble the CQWs into nanoscale patterns, EBL was used to pattern resist followed by the deposition of functionalized photosensitive NCs and UV exposure for ligand-replacement and lift-off. **Figure** 1a shows a schematic of the process flow. Figure 1b depicts photoluminescence (PL) spectrum of these CQWs along with their absorption spectrum at the emission peak centered at 647 nm with a FWHM of 25 nm.

As a suitable candidate to conduct measurements in visible light range, fused silica substrate was used on which the µ-laser was fabricated. A well-cleaned substrate was treated with 3-mercaptopropl-trimethoxysilane (MPTS) to aid in the adhesion of NCs to the substrate and to prevent delamination of the film.[33] After spin-coating of an electron beam resist on the prepared substrate to avoid charge accumulation during e-beam patterning, a conductive polymer film

(ESPACER) was deposited on the resist-coated substrate. The substrate was then patterned under electron beam exposure with EBL and developed to create the designed µ-laser structure. A detailed description of the fabrication steps can be found in the Methods section. A scheme of the µ-laser consisting of emitter and mirror is shown in Figure 1c.

Complete coverage of patterned trench structure with colloidal NCs is the next crucial step in such a way that voids and vacant sites between NCs were avoided. If not, lasing performance of the fabricated structure would be drastically affected. Indeed, when colloidal NCs are assembled into NC solids, the long, bulky, insulating ligands prevent close packing of these NCs and consequently undesired weak interparticle coupling adversely affects device performance. One of the promising solutions to increase the packing density and coupling strength of the NCs is to exchange the native capping ligand on the NC with smaller molecules either in solution and before film formation or by immersing their dried film in a solution containing smaller molecules.[34,35] As an alternative to these approaches, UV light was used to initiate ligand replacement. There have been several reports that have endorsed and demonstrated the capability of this technique. Jun *et al.* previously demonstrated the possibility of cross-linking the unsaturated double bonds between oleic acid ligands under UV illumination.[36] David *et al.* showed the decomposition of $CS_2N_3$ ligands to thiocyanate under UV exposure.[37] Kim *et al.* reported an approach to functionalize NCs by incorporation of the functional t-butoxycarbonyl (t-BOC) ligand which has an acid-labile moiety and activates upon UV exposure.[38] Recently, Talapin and collaborators have reported a comprehensive study on series of photochemically active surface ligands for NCs. They designed a broad class of photochemically active ligands using various photon energies including DUV, near-UV, blue, and visible light where the change in the surface chemistry was successfully exploited for direct patterning of NCs.[14] Although the direct lithography of NCs requires fewer steps with respect to conventional lithography, the reported patterns suffered from imprecise control of film thickness and remaining residual layer on the surface. However, in NC solution-based device fabrication process, fabrication of contamination-free, closed-packed, well-defined nanopatterns with controllable film thickness is essential, which remains a challenge to date.

As a novelty of this work, we have circumvented the aforementioned problems with combining electron beam lithography and UV-induced ligand exchange procedure. Here, µ-laser was first patterned with EBL and coated with NCs, then UV light exposure of the coated sample was ensure used to ligand exchange process, and finally a complete compact coverage of the design was obtained. Using this innovative approach, the resist can be patterned in nanoscale resolution and photoactive NC solution with compact photosensitive ligands can be used as building blocks of the desired patterned film. Lift-off process was performed by immersion of the sample in acetone/octane mixture under mild ultrasonic stirring. The end product of the mentioned steps is a robust, consistently crack/void-free film with strong adhesion to the surface. Atomic force micrograph (AFM) shows a uniform height of the deep-patterned CQW films with no chip-off after development of the pattern (Figure S1 in Supporting Information (SI)). High-resolution scanning electron microscopy of fabricated CQW self-resonant distributed Bragg reflector grating structure at different magnifications is presented in Figure 1d.

In the film formation step, the well-cleaned NCs capped with organic ligands such as oleic acid (OA) and oleylamine (OAm) were dispersed in octane and functionalized by incorporation of photo acid generating (PAG) surface ligands according to established procedures[13] (see Methods). Unlike regular methods, UV-assisted ligand exchange offers a substantial simplification of the processing steps. In this approach, the photochemically active solution is deposited on predefined patterns resist and let to dry. Subsequently, after UV exposure, PAG molecules undergo chemical transformations and generate strong acid. The acid interacts with the NCs surface and replaces lyophilic ligands with short moieties. Fourier transform infrared (FTIR) spectroscopy was used to monitor oleic acid left on the deep-patterned film. The exchange can be followed by comparing the vibrational spectrum before and after the exchange. **Figure** 2a shows diagnostic signals for native organic ligands at 2800−3000 cm$^{-1}$ for C−H stretching mode[39], which CQWs-OA thin films exhibited before UV exposure, were drastically suppressed after UV exposure. The new shorter ligands form an ion pair rather than a covalent bond with metal sites at the NC surface and lead to a substantial decrease in the interparticle separation. The comparison of TEM images in Figure 2b shows the synthesised CQWs with lyophilic ligands and large interparticle separation (2.15 nm) retained their size and morphology after UV-induced ligand exchange and the gaps between individual CQWs are reduced to 1.08 nm. Switching to shorter ligands has resulted in improved performance of solution-processed solar cells[40], photodetectors[41] and field-effect transistors (FETs)[42]. Similarly, the refractive index of the deep-patterned NC films plays a pivotal role in photonic applications. To determine the effective complex refractive index of the CQW films before and after UV-induced ligand exchange, variable-angle spectroscopic ellipsometry was used (see Methods). Unlike the intrinsic refractive index, the effective refractive index is sensitive to morphology, as it accounts for the NCs, ligands, and void space, which make up the overall film. The results in Figure 2c show that the films subjected to UV-induced ligand exchange procedure have a higher effective complex refractive index because the decreased bound ligand chain length during pattering process led to the increment of inorganic volume fraction, optical density and associated packing fraction of these films. The effective complex refractive index of NC solid films are dominated by the fill fraction of NCs, with only secondary influence from interparticle interaction and vary because of the particle size, the ligand chain length, and the deposition process.[43,44]

PL decay measurement was performed to estimate the decrease of PL quantum efficiency (PLQE) of the deep-patterned film of CQWs during patterning process with reference to their intact counterparts. Figure 2d shows the recorded PL decay curves for both the samples which were used to estimate the effect on the PLQE following the pattern formation and subsequent ligand exchange. The lifetime of the thin-film sample drops compared to the solution sample. However, there is a marginal difference before and after the patterning process of the thin film, albeit a slight decrement in the lifetime, suggesting insignificant modification of the PLQE (see Table S1). To check potential damage on the CQWs during the fabrication process, we compared QE of the CQW films before and after the fabrication process and found the deviation is in the range of a few percent (see Figure S2).

The design parameters of mirror and emitter slab thicknesses and lateral dimensions, waveguide width and emission wavelength of the µ-laser structure were carefully chosen to enable a low lasing threshold and efficient waveguide coupling via full electromagnetic solution using finite difference time-domain (FDTD) simulations. The resulting self-resonant CQW µ-laser attained by our direct deep-patterning approach forms an in-plane version of a Fabry–Pérot resonator consisting of DBR-based self-resonating gain section laterally sandwiched between a pair of DBR reflective regions. The gain part is composed of an array of deep-patterned CQW gratings with a film thickness of 300 nm. The mirrors are made of the same NCs, designed with a 200 nm pitch from multiple layers of alternating CQWs and air resulting in periodic variation in the effective refractive index and act as a high-quality reflector. The emitters are designed with a 500 nm pitch to be resonant at the emission wavelength of the CQW ~ 650 nm. To achieve a higher level of optical confinement, better beam quality and reduce the chance of optical damage to the emitting surface, the gain medium of the proposed µ-laser is modeled like multistripe index-guided laser arrays in a way that is surrounded by lower index $SiO_2$ (substrate) and air that behaves like a dielectric waveguide which ensures the confinement of photons to the active optical gain region and increases the rate of stimulated emission. However, the number and type of oscillating resonator modes in a laser depend on the circumstances under which a laser exhibits single-mode or multimode lasing. In this regard, two µ-laser structures with different lengths of gain medium (40 and 20 µm) were designed and lasing capabilities were evaluated.

To investigate lasing performance of the patterned structures, we conducted PL measurements where the patterned (40 µm in length) substrates were excited with a pulsed laser beam of 400 nm wavelength, ~110 fs pulse width and 1 kHz pulse rate (see Methods). The PL emission at different pump intensities was collected from the side of the substrate with an optical fiber. The resulting spectra of the deep-patterned CQWs are displayed in **Figure** 3a. For low pump intensities, a broad spectrum of spontaneous emission is observed, while narrow spectral features with FWHM around 0.6 nm appear above 21.0 µJ/cm$^2$. These features are attributed to the lasing modes due to their ultra-small FHWM accompanied with the superlinear increase of the intensity at the spectral peaks of these features beyond the threshold as shown in Figure 3b. After the appearance of the first lasing mode at ~662 nm, the second one emerges at ~650 nm with a threshold of 23.0 µJ/cm$^2$. By increasing the pump intensity even further beyond 100 µJ/cm$^2$, a third peak at ~638 nm is observed. The spectral spacing of these modes are about 12 nm, which agrees with the expected mode spacing of the designed CQW patterns. A possible approach to attain single-mode lasing is to reduce the length of cavity. However, this will inevitably increase lasing threshold of the µ-laser and require significantly high pumping intensity. Single-frequency operation is usually desirable and at the same time more difficult to achieve because it is not sufficient to introduce spatially varying loss or gain. Thereby, the same experiment was conducted using the shorter gain medium laser (20 μm in length). Figure 3c shows the pump intensity dependent spectra of the deep-patterned CQWs. We observe the lasing behavior at a threshold of 147 μJ/cm$^2$, which is evident from the spectral narrowing of the PL spectrum beyond the threshold as well as the superlinear increase of the intensity at the wavelength of interest as seen in Figure 3d.

Finally, to assess the reliability and feasibility of the proposed direct deep-patterning method in optoelectronic applications, our procedure is applied to different group of colloidal particles including colloidal quantum dots (QDs) and metal-oxide nanoparticles and also by adapting (CMOS)-compatible processes by changing electron beam lithography to photolithography. The mentioned nanoparticles are commonly used in pixelated lighting for liquid crystal displays. The high-resolution dual-color and multicolor lithographic patterning is also shown with our direct deep-patterning fabrication method. **Figure** 4a shows bicolor patterning of red-emitting CdSe/CdZnS CQWs and green-emitting CdSe/CdS QDs with electron beam lithography, which may provide a new and unique approach also for realizing efficient micro-LED displays. Ultrafine submicron-scale pixelated, high-efficiency, multicolor light sources integrated on a single chip are required by the display technologies of tomorrow. Figure 4b shows the bicolor patterning of two different NCs with photolithography (see Methods). High-quality patterns show capability of the method for creating multiple layers. For the second layer, the resist is directly spin-coated on top of the first patterned layer without any additional surface treatments. However, for a multilayer process, precise alignment is vital. The desired thickness of the NC layer is defined by resist thickness, which can be optimized by tuning the concentration of NCs in solution and multiplicity of the NC deposition. This technique is limited by the penetration depth of UV light in NCs to form thick patterns. To address this issue, NC deposition and UV exposure steps are repeated consecutively. Layer-by-layer spin-coating and ligand exchange are repeated, and after each exposure, substrate is immersed in octane to wash away the cleaved lyophilic ligands ensuring a complete exchange procedure. This modification enables shortening of the ligands, increasing the packing density and the refractive index of the resulting NC film. Figures 4c and 4d show high-quality, ultra-thick patterns of green-emitting QDs and red-emitting CQWs with high-aspect ratios, which are the highest values in their class reported to date, to the best of our knowledge. Fabrication of thick NC based Bragg gratings with small periods and high refractive index have so far not been reported. Figure 4e shows deep-patterned CQW bull's-eye geometry with a pitch of 650 nm. These bull's-eye geometries favor enhanced light extraction from light-emitting diodes and for demonstrating annular Bragg lasers. Figure 4f displays deep-patterned Au NPs. Detailed results are shown in Figure S3.

## 3. Conclusion

In conclusion, we have successfully developed a direct deep-nanopatterning technique based on EBL used in conjunction with UV-induced ligand exchange. This method allows to pattern semiconductor NCs and metal nanoparticles. Using EBL and photosensitive ligands, we have achieved nanoscale patterns with sub-100 nm resolutions while featuring heights as tall as 1.5 µm. Using this approach, we fabricated a high-quality self-resonant CQW µ-laser that enables tight confinement of light and thereby assures in-plane laser emission. The fabricated µ-laser is an excellent candidate to be used in integrated photonic circuits for practical applications such as lab-on-a-chip and optofluidics. This device is the first fully-colloidal self-resonant µ-laser, which operates with a low optical pump threshold of 21 µJ/cm$^2$ at room temperature.

## 4. Methods

*Materials:* A list of chemicals used and the commercial source for each are given in the (SI).

*Synthesis of red-emitting CdSe/CdZnS hot-injection shell-grown colloidal quantum wells (CQWs):* Initially, 4.5 monolayer thick CdSe core nanoplatelets were prepared using a well-known synthesis recipe reported in the literature.[31,45] CdSe/CdZnS CQWs were produced by using the recipe reported in our previous studies.[31,32,46,47] 1 mL of 4.5 ML CdSe core CQW in hexane having optical density of 2 (at 350 nm, using a 1 cm optical path length), Zn acetate (0.3 mmol), Cd acetate (0.1 mmol), ODE (10 mL) and OA (1 mL) were loaded in a three-neck quartz flask (50 mL). To remove impurities, oxygen and hexane from the solution, the reaction flask was kept under vacuum for 90 min at room temperature (RT), and then for 40 min at 85 °C. OLA (1 mL) was injected into the flask under argon-gas flow at 85 °C and the temperature of the system was increased to 300 °C. Shell growth was started by the injection at 160 °C using the ODE-octanethiol mixture that was prepared into the glovebox by mixing ODE (5 mL) and 1-octanethiol (140 µL). Injection rate was first set at 10 mL h$^{-1}$ until the temperature reached 240 °C, and then the rate was dropped to 4 mL h$^{-1}$ until all the precursor was injected into the reaction flask. The reaction was proceeded for an hour at 300 °C and then the reaction flask was cooled to RT by using water bath. Synthesized CQWs were separated from the solution by using the precipitation re-dispersion method. Finally, precipitated CQWs were re-dispersed in hexane (4 mL) and filtered for further usage.

*Synthesis of green-emitting CdSe/CdZnS quantum dots (QDs):* CdSe/ZnS QDs were prepared by using the previously reported synthesis recipe in the literature.[48-50] CdO (0.3 mmol), zinc acetate (4 mmol) and OA (5 mL) were loaded in a four-necked flask (50 mL). The mixture was heated to 150 °C under vacuum for 30 min. Then, the solution was cooled down and ODE (15 mL) was injected into the solution at 60 °C under argon-gas. After injection of ODE the solution is remain under vaccum for an extra hour. The temperature of the system was increased to 300 °C and the solution that contains cadmium-oleate and zinc-oleate became brightly transparent at 250 ºC. The precursor of TOP-Se-S was prepared into the glovebox by dissolving Se (0.3 mmol) and S (3 mmol) in TOP (2 mL) and it was rapidly injected into the solution at 300 ºC. Afterwards, the reaction was maintained at the same temperature for 10 min and then the flask was cooled down to RT using the water bath. 5 mL hexane was added into the as-synthesized QDs and QDs were precipitated by centrifugation using excess absolute ethanol. Finally, precipitated QDs were dissolved in hexane.

*Synthesis of Au nanoparticles (NPs):* Au NPs were synthesized according to the reported study in literature.[51] 20 mL OA and 1 mmol gold (III) chloride hydrate and (OA) were mixed into a three-neck flask and the solution was heated to 60° C for half an hour under argon gas flow. Then, the temperature was increased to 150 °C and kept for 2 hours. Afterward, the flask was cooled down to RT and the solution was precipitated by methanol. Eventually, precipitated gold NP was dispersed in hexane and filtered before using.

*μ-Laser device fabrication:* The substrate was washed in an ultrasonic bath, with a sequence of solvents as follows: alkaline detergent, deionized water, acetone, ethanol (EtOH) and deionized water. The dry substrate was immersed in a 2% v/v solution of mercaptopropyltrimethoxysilane (MPTS) in acetone for 30 min in a sealed PVC vessel. Coated substrate was then rinsed by acetone and acetone/ethanol. All substrate preparation was carried out in air and room temperature. Then, 300 nm thick layer of positive resist such as polymethylmethacrylate (PMMA-A4) was spin-coated for EBL on the prepared substrate at 1,500 rpm for 60 s. The resist was baked at 180 °C for 300 s and followed by deposition of a 20 nm ESPACER. The e-beam exposure was carried out at a 100 μC/cm$^2$ dose range, and the sample was developed using methyl isobutyl ketone: isopropyl alcohol (MIBK: IPA) with a1:2 ratio for 45 s and followed by 15 s fresh IPA and then 30 s in DI water.

The NCs were washed twice with EtOH and dispersed in octane with a concentration of 65 mg/mL and functionalized by incorporation of $(C_6H_5)_3S^+OTf^-$, (triphenyl sulfonium triflate) PAG molecules. PAG molecules were dissolved in EtOH and mixed with the NC solution (5%, by weight). The NCs was spin-coated on the predefined patterned resist and dried for 20 min to evaporate the solvent and irradiated under a 254-nm DUV lamp with a dosage of 250 mJ/cm$^2$. Under UV exposure, $(C_6H_5)_3S^+OTf^-$ molecules decompose and generate a strong acid, triflic acid (HOTf). The strong acid interacts with the NCs surface and the photogenerated protons efficiently attack lyophilic ligands and replace them with OTf groups. As a very weak nucleophile, triflate forms an ion pair rather than a covalent bond with metal sites at the NC surface. The new OTf$^-$ ligands represent a substantial decrease in ligand size [13]. The exposed samples were immersed in acetone/octane mixture under mild ultrasonic stirring for time varied from 5 to 15 s depending on the resist thickness and pattern resolution.

*Lasing experiments:* A laser amplifier (Spitfire Pro, Spectra Physics) at 800 nm wavelength, 110 fs pulse width and 1 kHz pulse rate were used as the excitation source. The wavelength was converted to 400 nm by a frequency-doubling BBO crystal. The excitation intensity was adjusted with a variable neutral density filter. The 400 nm beam was focused onto the substrate with the help of a convex cylindrical lens and passed through a slit. The PL emission of the CQW patterns was collected from the side with the aid of an optical fiber. The spectra at different pump intensities were recorded with an optical spectrometer (Avantes).

*Optical and electron beam lithography:* For thick patterns SML1000 resist was spin-coated for EBL on the prepared substrate at 1,800 rpm for 60 s using supplier-recommended conditions (1.5 μm thickness /1,800 rpm). A positive photo resist like AZ 5214 and AZ 400K development were used in photolithography step and after patterning of the first NC layer film, the second layer was directly patterned on top of the first.

*Characterization techniques:* Transmission electron microscopy (TEM) was performed using JEOL 2100F electron microscope. Scanning electron microscopy (SEM) was carried out on a FEI NovaLab 600i SEM/ FIB instrument operating at 15 kV. UV-vis absorption and PL spectra were collected using a Agilent Varian Cary 5000 UV-Vis-NIR and Cary Eclipse spectrophotometer. Atomic force microscopy (AFM) was performed using Asylum Research MFP-3D instrument. Fourier transform infrared (FTIR) spectra were acquired in transmission mode using a Bruker Vertex 70 FTIR spectrometer. Ellipsometry measurements were performed by using a J.A. Woollam Co., Inc variable angle spectroscopic ellipsometer (VASE). For spectroscopic ellipsometry measurements, data was collected at angles between 65-70° and 75° over a wavelength range 500−1200 nm. To model the spectroscopic ellipsometry data and extract the effective complex refractive index, the spectral data was split into two regimes. At low energies, where absorption is negligible (750−1000 nm), a Cauchy fit was initially used. This allowed the CQW film thickness, with an initial guess given by AFM or profilometrydata, to be refined as the imaginary portion of the permittivity was negligible in the region. After the transparent region was fit, the CQW film thickness was fixed and the wavelength range was slowly expanded so that spectral features above the band gap energy could be fit using Kramers−Kronig consistent dispersion models. Gaussian oscillators were chosen to describe the absorption features in the visible region, as they better describe the slight polydispersity present in the CQW ensemble and any resulting inhomogeneous line-broadening when compared to Lorentz oscillators. The final dispersion model for the CQW film consisted of the sum of five individual oscillators. The acquired data analysis were carried out using W-Vase32 software package. For this dispersion model, the resulting meansquared error (MSE) of the standard deviation between the measured and generated reflectivity data was 1.759. Time resolved fluorescence (TRF) measurements were conducted using FluoTime 200 time-resolved Spectrometer. An ultra violet laser (375 nm wavelength) with a pulse width of 200 ps and pulse repetition of 2.5 MHz was used to excite our sample. We acquired emitter lifetime by means of the time-correlated photon counting utilizing Pico Harp time correlator with a temporal resolution of 4 ps at the emitter peak wavelength of 652 nm. Then, we fitted the decay curves with multiexponential decays $\sum_i A_i e^{-t/\tau_i}$ after convolving with the impulse response function. The amplitude average lifetimes were determined as $\tau_{avg} = \sum_i A_i e^{-t/\tau_i} / \sum_i A_i$.

**Supporting Information**
Supporting Information is available from the Wiley Online Library or from the author.

**Acknowledgements**

The authors acknowledge the financial support from TUBITAK through 115E679, 115F297 and 117E713 programs. The authors thank Mr. Mustafa Guler and Mr. Ovunc Karakurt for their assistance in TEM imaging, Dr. Gokce Celik for her help on the ellipsometric measurements and Mr. Semih Bozkurt for his help on the AFM characterization.

Received: ((will be filled in by the editorial staff))
Revised: ((will be filled in by the editorial staff))
Published online: ((will be filled in by the editorial staff))

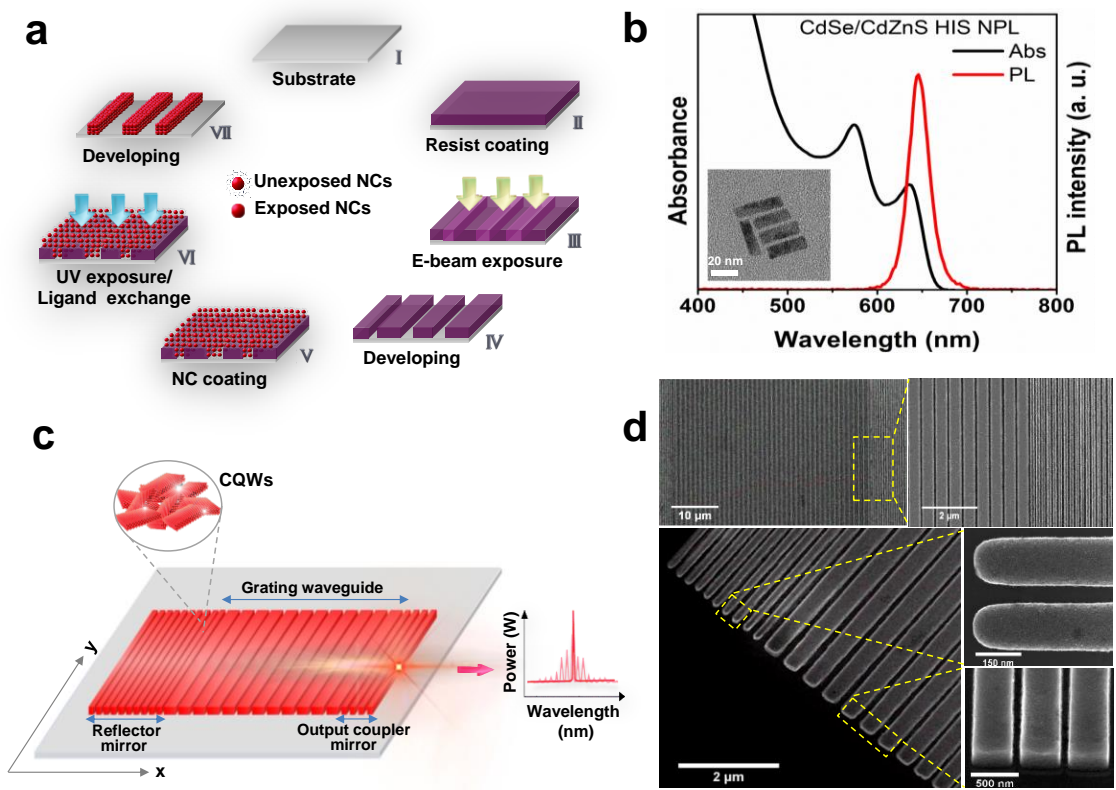

**Figure 1.** (a) Flow of the processing steps for fabricating patterned NC films. (b) PL and absorption spectra of CdSe/Cd$_{0.25}$Zn$_{0.75}$S core/HIS CQWs. The inset shows high-resolution transmission electron microscopy (HR-TEM) image of the CQWs (the scale bar is 20 nm). (c) Schematic view of the designed self-resonant all-colloidal laser made of DBR waveguides and mirrors. (d) Scanning electron micrograph of the final colloidal device structure at different magnifications. (Mirrors and emitters have 200 and 500 nm pitches, respectively).

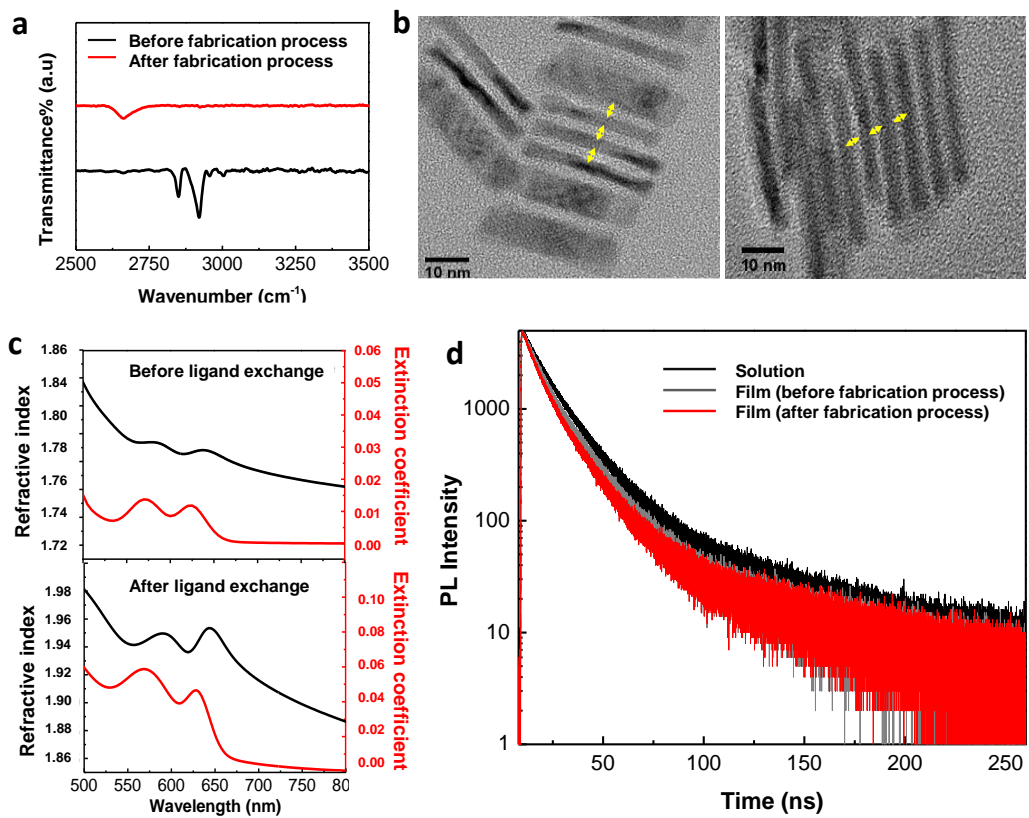

**Figure 2.** (a) FTIR spectra of CQWs capped with native oleate ligands and after UV exposure on Si substrate. (b) TEM images of CQWs before and after UV exposure. (c) Real and imaginary effective refractive indices of the CQW thin film before and after ligand exchange, measured via spectroscopic ellipsometry. (d) PL decay kinetics of CQW ensembles before and after the patterning process.

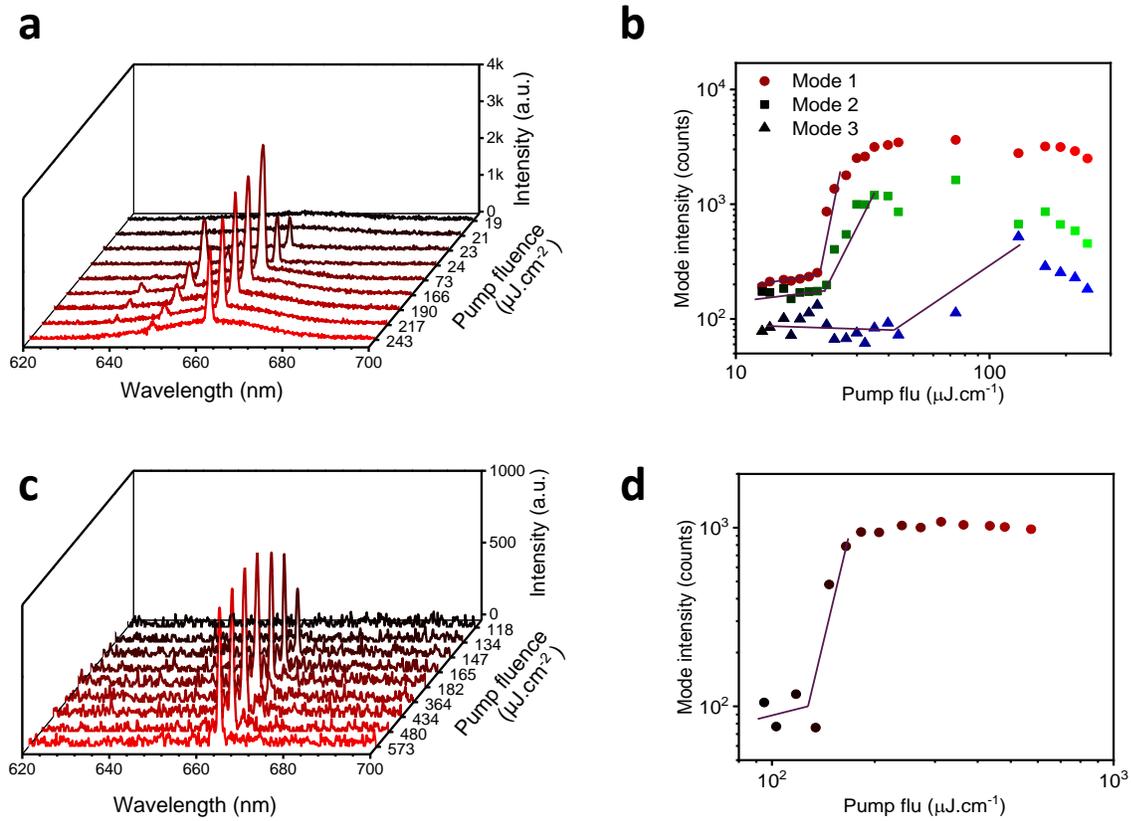

**Figure 3.** (a) PL spectra of the deep-patterned CQWs under various excitation intensities. Three lasing modes at 662, 650 and 638 nm appear with onsets of ~21, ~23 and ~130 μJ/cm$^2$, respectively. (b) Peak intensities of the three lasing modes at ~662 (blue up-triangles), ~650 (red squares) and ~638 nm (orange down-triangles). (c) PL spectra of the deep-patterned CQWs under pulsed excitation of various intensities. (d) Peak intensity of the spectra as a function of the pump fluence. The steep increase in the curve corresponds to the onset of the lasing at a threshold of ~140 μJ/cm$^2$, followed by the saturation of the peak intensity.

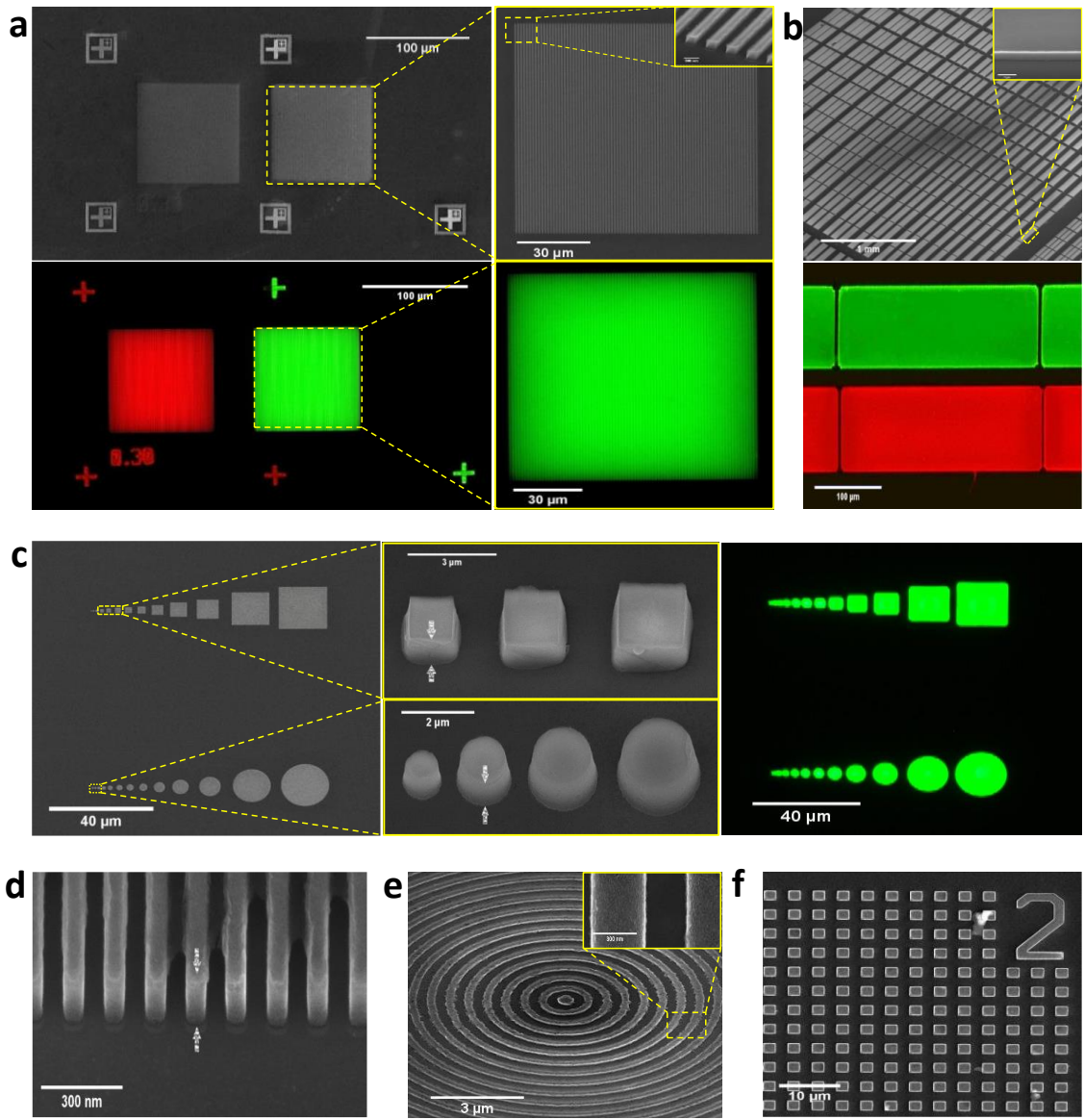

**Figure 4.** Bilayer of red-emitting CdSe/CdZnS CQWs and green-emitting CdSe/ZnS QDs patterned with (a) EBL (the scale bar in inset is 500 nm), (b) photolithography (the scale bar in inset is 1 μm) and imaged with SEM and PL microscopy. (c) High-quality and thick patterned green-emitting CdSe/ZnS QDs imaged with SEM and PL microscopy, (d) red-emitting CdSe/CdZnS CQWs, (e) Bragg gratings patterns of red-emitting CdSe/CdZnS CQWs and (f) deep-patterned Au NPs imaged with SEM.